\RequirePackage[hyphens]{url}
\documentclass[aps,pre,twocolumn,superscriptaddress]{revtex4-2}

\usepackage{natbib}
\usepackage{hyperref}
\hypersetup{colorlinks = true, urlcolor = blue,	linkcolor = blue, citecolor = blue}

\usepackage{xcolor}
\usepackage{graphicx}
\usepackage{siunitx}
\usepackage{amsmath,amssymb}

\begin{document}

\title{Self-sustained patchy turbulence in shear-thinning active fluids}

\author{Henning Reinken}
\email{henning.reinken@ovgu.de}
\affiliation{Institut f\"ur Physik, Otto-von-Guericke-Universität Magdeburg, Universitätsplatz 2, 39106 Magdeburg, Germany} 

\author{Andreas M. Menzel}
\email{a.menzel@ovgu.de}
\affiliation{Institut f\"ur Physik, Otto-von-Guericke-Universität Magdeburg, Universitätsplatz 2, 39106 Magdeburg, Germany}

\date{\today}

\begin{abstract}
Bacterial suspensions and other active fluids are known to develop highly dynamical vortex states, denoted as active or mesoscale turbulence. 
We reveal the pronounced effect of non-Newtonian rheological conditions on these turbulent states, concentrating on shear thinning.
A self-sustained heterogeneous state of coexisting turbulent and quiescent areas develops, which results in anomalous velocity statistics.
The heterogeneous state emerges in a hysteretic transition when varying activity.
We provide an extensive numerical analysis and observe features consistent with a directed percolation transition. 
Our results are important, for instance, when addressing active objects in biological media with complex rheological properties. 
\end{abstract}

\maketitle

\noindent{\large\textbf{Introduction}}\\[-0.5\baselineskip]

\noindent Bacterial suspensions are a paradigmatic example of active matter~\cite{vicsek1995novel,marchetti2013hydrodynamics,Bechinger2016}.
Due to the constant energy input on the scale of bacterial microswimmers, such suspensions exhibit various kinds of spatiotemporal dynamics~\cite{jeckel2019learning}.
These include biofilm formation~\cite{hall2004bacterial} as well as collective motion and swarming states~\cite{be2019statistical,be2020phase}, which enable the rapid expansion to new territories.
In particular, swimming bacteria display swirling and vortex patterns~\cite{dombrowski2004self,sokolov2007concentration,sokolov2012physical,wensink2012meso}, which have been denoted as active or mesoscale turbulence~\cite{wensink2012meso,alert2021active}.
In contrast to inertial turbulence, these states exhibit a characteristic vortex size.
observed, for example, in suspensions of \textit{Bacillus subtilis}~\cite{dombrowski2004self,wensink2012meso,nishiguchi2018engineering}.
The key features are captured by continuum-theoretical descriptions \cite{wensink2012meso,dunkel2013minimal,slomka2015generalized,reinken2018derivation} in terms of the velocity field. 

Both the propulsion mechanism and the interactions between microswimmers are mediated by the solvent medium.
In general, the environments inhabited by bacteria and other biological microswimmers often display non-Newtonian and viscoelastic properties~\cite{li2021microswimming}.
Examples include spermatozoa in the reproductive tract~\cite{fauci2006biofluidmechanics} and pathogenic bacteria in gastric mucus 
or other extracellular fluids~\cite{celli2009helicobacter}.
Predominantly, blood shows shear thinning and can display complex rheological properties such as viscoelasticity and thixotropy~\cite{beris2021recent}.
During biofilm formation, bacteria excrete extracellular polymeric substances~\cite{hall2004bacterial,worlitzer2022biophysical}, likewise resulting in non-Newtonian rheology~\cite{jana2020nonlinear}.

The development of realistic models for collective motion of microswimmers thus necessitates to include non-Newtonian effects.
So far, complex solvents have mostly been investigated in the case of single-swimmer dynamics, for instance, exploring the impact of viscoelasticity and shear thinning or thickening on the swimming speed~\cite{teran2010viscoelastic,li2015undulatory,datt2015squirming}.
These studies show that, depending on various parameters such as swimmer geometry and fluid properties, complex rheology can enhance or hinder self-propulsion~\cite{montenegro2013physics,van2022effect}. It allows for reciprocal deformations of the swimmers to achieve propulsion~\cite{qiu2014swimming,datt2018two,yasuda2020reciprocal,eberhard2023reciprocal}, in contrast to Newtonian fluids featuring time-reversible Stokes flow~\cite{purcell1977life,lauga2009hydrodynamics}.

So far, only a limited number of studies explore the impact of complex rheological properties on the collective motion of microswimmers.
Here, the focus has generally been on viscoelastic fluids. Diverse spatiotemporal pattern formation is observed, which results in enhanced complexity or calming effects depending on the elastic parameters~\cite{bozorgi2013role,bozorgi2014effects,hemingway2015active,hemingway2016viscoelastic,li2016collective,plan2020active}.
Recent studies have further demonstrated that complex rheological properties such as shear thickening~\cite{reinken2024vortex} and viscoelasticity~\cite{liu2021viscoelastic} can be utilized to control emergent states.
However, in experiments bacterial suspensions often display shear thinning due to the presence of extracellular polymers~\cite{jana2020nonlinear}. The impact of such shear-thinning rheological conditions on active turbulence are still to be explored. 

We address this open question through a recent continuum theory~\cite{slomka2015generalized,slomka2017geometry,slomka2017spontaneous}
that shows turbulent dynamics consistent with experimental observations on both bacterial microswimmers and ATP-driven microtubular networks~\cite{slomka2017spontaneous}.
Its versatility, besides characterizing pure active turbulence, has been demonstrated, for example, by extensions for shear-thickening active suspensions~\cite{reinken2024vortex}.
We now provide the missing and significantly more abundant case of mesoscale turbulence in shear-thinning suspensions. That is, viscosity locally decreases with increasing local shear rate. 
We explore the resulting patterns of mesoscale turbulence and find, as a key observation, that shear thinning leads to hysteretic behavior of the turbulent state when varying the activity. More precisely, a regime of coexisting patterns of turbulence and macroscopically quiescent patches (vanishing collective velocity) emerges through shear thinning. 
Anomalous velocity statistics are found in this regime.
Thus, we uncover a self-sustained dynamic state of heterogeneity combining regions of turbulence with nonturbulent patches.\\

\noindent{\large\textbf{Results}}\\[-0.5\baselineskip]

\noindent {\textbf{Mesoscale turbulence}}

\begin{figure*}
\includegraphics[width=0.999\linewidth]{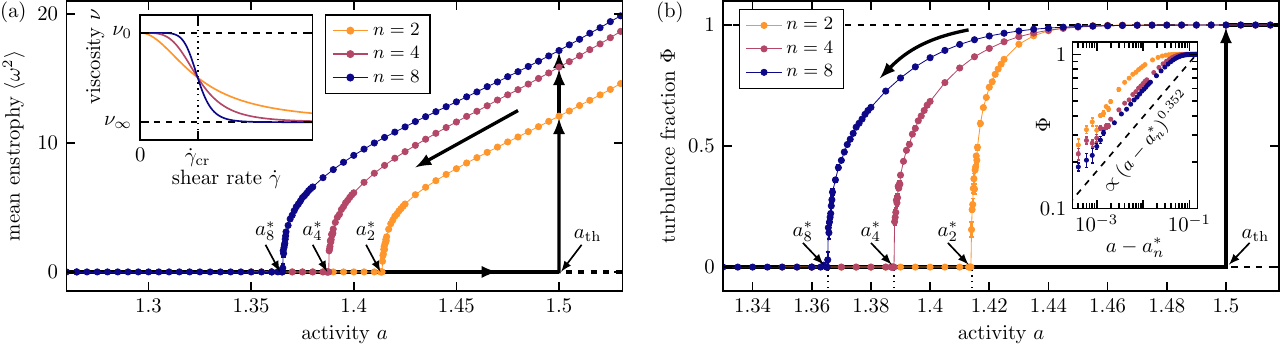}
\caption{\label{fig:transition}Hysteretic behavior. Hysteresis loops for the degree of mesoscale turbulence as a function of microswimmer activity $a$ for parameters of shear thinning $\nu_0/\nu_\infty = 1.5$, $\zeta = 2$, and different $n$. For increasing $a$, the isotropic, macroscopically quiescent state becomes linearly unstable beyond $a_\mathrm{th}$. When decreasing $a$, turbulence persists down to the critical values $a^*_n<a_\mathrm{th}$. (a) Quantification via the system-averaged enstrophy $\langle \omega^2 \rangle$. Inset: considered effect of shear thinning as described by the dependence of the local viscosity $\nu$ on the local shear rate $\dot{\gamma}$ in the framework of the Cross fluid model~ \cite{cross1965rheology,barnes1989introduction}.
(b) Quantification via the time-averaged area fraction of turbulent patches $\Phi$. Inset: $\Phi$ as a function of the distance to the critical value, $a-a_n^\ast$, revealing a regime of power-law scaling. Error bars denote the standard error.} 
\end{figure*}

\noindent To quantify the given situation, we consider a recent continuum theory~\cite{slomka2015generalized,slomka2017geometry,slomka2017spontaneous,reinken2024vortex} that describes the dynamics of the overall (collective) velocity field $\mathbf{v}(\mathbf{x},t)$ of the entire incompressible suspension via a generalized Navier--Stokes equation,
\begin{equation}
\label{eq:NSt}
\partial_t \mathbf{v} + \mathbf{v} \cdot \nabla \mathbf{v} = - \nabla \tilde{p} + \nabla \cdot \tilde{\boldsymbol{\sigma}}\, , \qquad \nabla \cdot \mathbf{v} = 0\, .
\end{equation}
Here, the constant density $\rho$ is absorbed into the pressure and stress tensor, $\tilde{p} = p/\rho$ and $\tilde{\boldsymbol{\sigma}} = \boldsymbol{\sigma}/\rho$.
The stress tensor is expanded in gradients of the deformation rate tensor $\boldsymbol{\Sigma} = [(\nabla \mathbf{v}) + (\nabla \mathbf{v})^\top]/2$, 
\begin{equation}
\label{eq:stressExpansion}
\tilde{\boldsymbol{\sigma}} = (\Gamma_0 + \Gamma_2 \nabla^2 + \Gamma_4 \nabla^4 ) \big[ (\nabla \mathbf{v}) + (\nabla \mathbf{v})^\top \big]\, ,
\end{equation}
where $^\top$ marks the transpose\cite{slomka2015generalized,slomka2017geometry,slomka2017spontaneous,reinken2024vortex}. 
Key features are the emergence of highly dynamic vortex patterns and the selection of a specific length scale of the vortex size.
In the context of microswimmer suspensions, the resulting state is usually denoted as mesoscale turbulence~\cite{wensink2012meso}.

For $\Gamma_2 = \Gamma_4 =0$, the theory reduces to the familiar case of vanishing activity.
Then, $\Gamma_0$ corresponds to the kinematic viscosity ${\nu}$.
We adopt the relation $\Gamma_0 = {\nu}$ for the active case.
The coefficient $\Gamma_4$ must be positive to ensure stability at short wavelengths.
Activity of the swimmers increases the coefficient $\Gamma_2$, which excites patterns of intermediate wavelengths once $\Gamma_2 > \sqrt{4 {\nu} \Gamma_4}$.
A linear stability analysis reveals a critical finite wavenumber $k_\mathrm{c} = \sqrt{\Gamma_2/2\Gamma_4}$.
The resulting band of unstable modes indicates wavenumbers at which energy is  pumped into the system as a result of the intrinsic activity of the microswimmers.
Subsequently, nonlinear advection, represented by $\mathbf{v}\cdot \nabla \mathbf{v}$ in Eq.~\eqref{eq:NSt}, is responsible for the development of turbulence and associated energy transport between wavenumbers.
The resulting balance between energy input and dissipation leads to a statistically stationary state~\cite{slomka2015generalized,slomka2017geometry,slomka2017spontaneous,reinken2024vortex}.

For further details of the theoretical description, we refer to Supplementary Note 1.
There, we also point out important differences to the Toner--Tu~\cite{toner2005hydrodynamics} and Toner--Tu--Swift--Hohenberg equations~\cite{wensink2012meso,reinken2018derivation,alert2021active}.
The latter represent a related framework to model mesoscale turbulence.
Compared to those, the here-employed description is formulated in terms of the overall velocity of the suspension instead of the velocity of only the microswimmers.
Consequently, incorporating non-Newtonian rheology is straightforward.

To investigate how the interplay of non-Newtonian rheology and active energy input impacts the emerging dynamic structures,
we now turn to the abundant case of shear-thinning active suspensions.
In a minimal approach, the dependence of the local viscosity ${\nu}(\mathbf{x})$ on the local shear rate $\dot{\gamma}(\mathbf{x})$ is described in terms of the frequently considered Cross fluid~\cite{cross1965rheology,barnes1989introduction}.
Its main feature is a viscosity ${\nu}_\infty$ at high shear rate that is smaller than the viscosity ${\nu}_0$ at low shear rate, accompanied by a continuous crossover region in between, 
\begin{equation}
\label{eq:viscosity}
{\nu}(\mathbf{x}) = {\nu}_\infty + \frac{{\nu}_0 - {\nu}_\infty}{1 + [\dot{\gamma}(\mathbf{x})/{\dot{\gamma}_\mathrm{cr}}]^n}\, .
\end{equation}
The local shear rate is calculated from the deformation rate tensor via $\dot{\gamma}(\mathbf{x}) = \sqrt{2\boldsymbol{\Sigma}(\mathbf{x}):\boldsymbol{\Sigma}(\mathbf{x})}$.
While the exponent $n$ quantifies the steepness of the crossover, the reference shear rate $\dot{\gamma}_\mathrm{cr}$ locates this crossover, see the inset of Fig.~\ref{fig:transition}(a). 
In our case, the non-Newtonian properties included in our model are induced by the carrier fluid and thus appear in the continuum description, which represents both microswimmers and solvent by one velocity field.
Bacterial suspensions during biofilm formation provide an example of this situation.
There, extracellular polymers excreted by the bacteria lead to complex, non-Newtonian rheology of the suspension~\cite{jana2020nonlinear}.

\begin{figure*}
\includegraphics[width=0.999\linewidth]{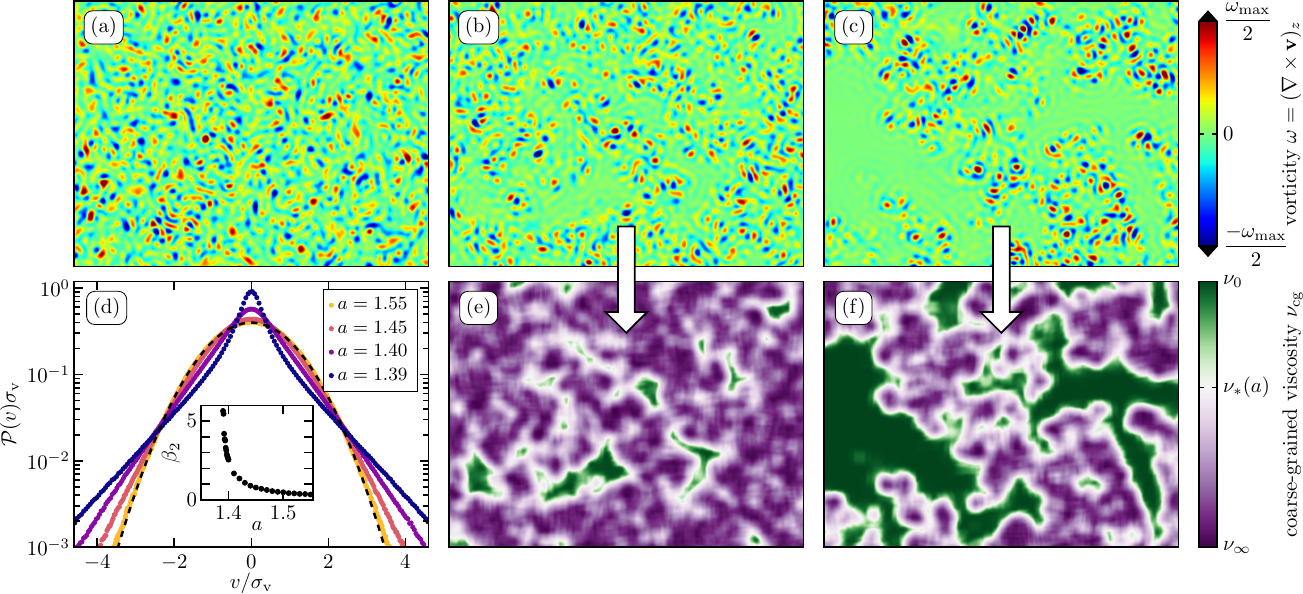}
\caption{\label{fig:snapshots}Coexistence of turbulent and macroscopically quiescent regions. Snapshots of the vorticity field $\omega$ (rescaled by its respective maximum value $\omega_\mathrm{max}$) for activities (a) $a = 1.55>a_{\mathrm{th}}$, (b) $a = 1.41<a_{\mathrm{th}}$, and (c) $a=1.39$. Quiescent patches of increasing size develop for decreasing $a$. 
(d) Velocity distribution function $\mathcal{P}(v)$.
Statistics are very close to Gaussian (indicated by the dashed line) for large activity $a$. Yet, they develop pronounced tails and an elevated maximum for decreasing $a$.
Both axes are normalized using the standard deviation $\sigma_v$.
Inset: excess kurtosis $\beta_2$ as a function of $a$, quantifying deviations from Gaussian statistics.
(e,f) Snapshots of the locally averaged, coarse-grained viscosity $\nu_\mathrm{cg}(\mathbf{x})$ for the same values of $a$ as in (b,c). Remaining parameters are $\nu_0/\nu_\infty = 1.5$, $n=4$, $\zeta = 2$, and the size of the snapshots is $64\pi \times 48\pi$.}
\end{figure*}

We utilize the length $k_\mathrm{c}^{-1}$, the time $(k_\mathrm{c}^2 {\nu}_\infty)^{-1}$, and the speed $k_\mathrm{c} {\nu}_\infty$ to rescale Eq.~(\ref{eq:NSt}),
\begin{equation}
\label{eq:generalizedNStRescaled}
\partial_t \mathbf{v} + \mathbf{v} \cdot \nabla \mathbf{v} = - \nabla \tilde{p} + \nabla \cdot (2\nu \boldsymbol{\Sigma}/\nu_\infty)  + a (2\nabla^4 \mathbf{v} + \nabla^6 \mathbf{v})\, ,
\end{equation}
where $\nu/\nu_\infty = 1 + (\nu_0/\nu_\infty - 1)/\{1 + [\sqrt{2\boldsymbol{\Sigma}:\boldsymbol{\Sigma}}/\zeta]^n\}$.
Thus, the parameters characterizing shear thinning are the viscosity ratio $\nu_0/\nu_\infty$, the rescaled crossover shear rate $\zeta = \dot{\gamma}_\mathrm{cr}/(k_\mathrm{c}^2 \nu_\infty)$, and the exponent $n$.
Moreover, $a = \Gamma_2^2/(4 \nu_\infty \Gamma_4)$ sets the strength of the activity of the microswimmers.
Finally, $\mathbf{v}(\mathbf{x},t)$ quantifies the solvent ﬂow on length scales several times larger than that of individual microswimmers.
Thus, the isotropic state $\mathbf{v}(\mathbf{x},t) = \mathbf{0}$ describes macroscopic quiescence.
In this state, the swimmers are active on the microscopic scale, but their orientations are disordered, resulting in the absence of collective motion.

A straightforward stability analysis shows that the quiescent state $\mathbf{v}(\mathbf{x},t) = \mathbf{0}$ is linearly stable below a threshold activity $a_\mathrm{th} = \nu_0/\nu_\infty$ (see Methods).
Intuitively, viscosity counteracts turbulence. 
Thus, the active driving has to overcome the hindrance by the viscosity $\nu_0$ for turbulent instabilities to set in.
Increasing the activity above $a_\mathrm{th}$, this state becomes unstable with respect to the growth of turbulent patterns characterized by a band of unstable modes.
The fastest-growing mode $k_\mathrm{m}$ is close to the critical mode $k_\mathrm{c}=1$ for values of $a$ close to $a_\mathrm{th}$.
For $a = 0$, we recover the familiar Navier--Stokes equation without activity for a shear-thinning Cross fluid.\\

\noindent \textbf{Emerging heterogeneous patterns}

\noindent To analyze the emerging patterns that result from this combination of mesoscale turbulence and shear thinning, we solve Eq.~(\ref{eq:generalizedNStRescaled}) numerically for a large two-dimensional system of size $128\pi \times 128\pi$ with periodic boundary conditions (see Methods).
We start from random initial conditions and set the parameters to $\zeta = 2$ and $\nu_0/\nu_\infty = 1.5$, while varying the activity $a$.
The viscosity ratio is in the range of recent experimental results on bacterial biofilms, where the rheological properties of the extracellular polymer matrix resulted in a viscosity decrease at high shear rates of up to $\SI{75}{\percent}$ depending on the species~\cite{jana2020nonlinear}.
We begin below the threshold, $a<a_{\rm{th}}$,
and then increase activity $a$. Turbulence develops across the whole system once the threshold $a_{\mathrm{th}}$ is passed, where the quiescent state becomes linearly unstable, see the vertical black arrow in Fig.~\ref{fig:transition}(a). There, the degree of mesoscale turbulence is quantified by the mean enstrophy $\langle \omega^2 \rangle$, where we define the vorticity field $\omega = (\nabla \times \mathbf{v})_z$. 

The case of decreasing activity $a$ features even more interesting behavior. We start above the threshold, $a>a_{\textrm{th}}$, implying that the macroscopically quiescent state is linearly unstable. Thus, vortex patterns grow across the whole system according to the finite-wavelength instability discussed above.
For illustration, Fig.~\ref{fig:snapshots}(a) shows a snapshot of the vorticity field at $a = 1.55 > a_{\textrm{th}}$.
When now decreasing the activity $a$, turbulence persists down to below the threshold value $a_\mathrm{th}$.
We have not observed such hysteretic behavior for regular Newtonian fluids.
The associated hysteresis loops of mesoscale turbulence, quantified by the mean enstrophy $\langle \omega^2 \rangle$, are plotted in Fig.~\ref{fig:transition}(b) for different values of $n$.
To explain this hysteretic behavior, that is, the persistence of existing turbulence when decreasing the activity down to $a<a_{\mathrm{th}}$, we note the following. 
The actively induced turbulent state is characterize by relatively high local shear rates. Locally, due to shear thinning, these high shear rates significantly reduce viscosity.
Reduced viscosity, in turn, favors turbulence. 
This feedback mechanism thus provides a channel for the self-sustenance of turbulence below the instability threshold $a_\mathrm{th}$, resulting in hysteresis.

Decreasing activity further, we observe increasingly heterogeneous states, see the snapshots in Figs.~\ref{fig:snapshots}(b) and~(c). Compared to Fig.~\ref{fig:snapshots}(a), the system now exhibits turbulent regions coexisting with increasingly large quiescent areas devoid of vortices.
This emergent heterogeneous state is highly dynamic and the locations of the macroscopically  quiescent patches continuously change while the vortex patterns rearrange (see also Supplementary Movie 1).

The development of heterogeneous states of coexistence also significantly changes the velocity statistics $\mathcal{P}(v)$. Here, $v$ represents an arbitrary component of the velocity field, $v=v_x$ or $v=v_y$. Due to macroscopic isotropy, see Supplementary Note 2 and Supplementary Figure 1, either component can be used.
In previous studies, employing a similar statistical evaluation~\cite{wensink2012meso,bratanov2015new,james2018vortex,mukherjee2023intermittency} for the Newtonian case, the velocity statistics were found to be very close to Gaussian
in mildly active regimes.
Strong activity, however, may lead to anomalous statistics~\cite{mukherjee2023intermittency}.
In our case, we observe Gaussian behavior in the fully turbulent state for $a>a_{\mathrm{th}}$, see Fig.~\ref{fig:snapshots}(d), suggesting a mildly active regime.
However, when decreasing activity below the threshold value $a_{\mathrm{th}}$, the distribution function develops pronounced tails together with a sharper maximum at $v=0$. 
In our situation of shear thinning, these features reflect heterogeneous states of coexisting macroscopically quiescent areas ($v\approx 0$) and turbulent regions of elevated local macroscopic velocities.

Since already the turbulent regions themselves are very heterogeneous, we employ a coarse-graining procedure to further quantify the emerging structures.
First, we determine the local viscosity from local shear rates via the rescaled version of Eq.~(\ref{eq:viscosity}).
Then, across the system, we locally average the viscosity over square regions of a size corresponding to the critical length scale $2\pi k_c^{-1}$ (see Methods).
Snapshots of the resulting coarse-grained viscosity field $\nu_\mathrm{cg}(\mathbf{x})$ are depicted in Figs.~\ref{fig:snapshots}(e) and (f). To distinguish between turbulent and macroscopically quiescent regions, we introduce a color scheme. If the locally coarse-grained viscosity is large enough to suppress the linear instability and thus turbulence, $\nu_\mathrm{cg}(\mathbf{x}) > \nu_\ast(a) = a\,\nu_\infty$ (see Methods), we employ green color. 
Contrarily, low-viscosity regions of locally self-sustained turbulence are dyed in purple. 
In particular, Figs.~\ref{fig:snapshots}(e) and (f) demonstrate the coexistence of turbulent and macroscopically quiescent regions of comparable area fraction.
Further visualizations are provided in Supplementary Figure 2 and Notes 3 and 4. 
(Supplementary Movie 1 shows the stable coexistence of rearranging domains at a constant activity, whereas Supplementary Movie 2 shows how the turbulent domains grow when the activity is increased suddenly from a value in the regime of coexistence.) \\

\begin{figure}
\includegraphics[width=0.99\linewidth]{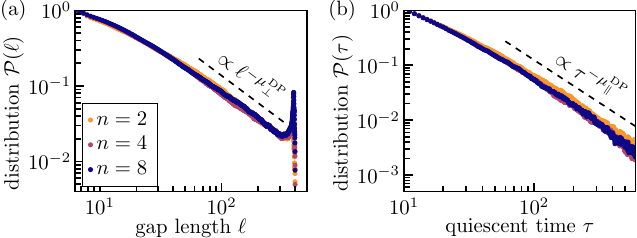}
\caption{\label{fig:DP}Structures at criticality. Features of the spatial and temporal structures very close to the critical point in terms of the distributions $\mathcal{P}$ of (a) the distances (lengths of the gaps) between the turbulent regions $\ell$ and (b) the time intervals of local quiescence $\tau$.
The distributions are normalized by the value at smallest $\ell$ or $\tau$.
The parameters are $\zeta = 2$, $\nu_0 = 1.5 \nu_\infty$, and activity is $a=1.41460$ for $n=2$, $a=1.38825$ for $n=4$, and $a=1.36585$ for $n=8$. 
The activities are chosen as close to the critical points as possible, compare Fig.~\ref{fig:transition}.
An emergent power-law scaling near the transition for elevated $\ell$ and $\tau$ is consistent with $2+1$ directed percolation (``DP''), as indicated by the dashed lines, which represent the exponents of directed percolation $\mu_\perp^\mathrm{DP} = 1.204(2)$ and $\mu_\parallel^\mathrm{DP} = 1.5495(10)$~\cite{munoz1999avalanche,takeuchi2007directed}.}
\end{figure}

\noindent \textbf{Critical behavior at the transition to turbulence}

\noindent To further quantify the transition from fully developed turbulence to macroscopically quiescent states, we determine the time-averaged area fraction of turbulence $\Phi$ via the coarse-grained viscosity.
We define the fluid at position $\mathbf{x}$ as turbulent if $\nu_\mathrm{cg}(\mathbf{x}) < \nu_*(a)= a\, \nu_\infty$.
$\Phi$ acts as an order parameter. Its bounds indicate a macroscopically quiescent state of the entire system ($\Phi = 0$) and fully turbulent ones ($\Phi = 1$).
Figure~\ref{fig:transition}(b) shows $\Phi$ as a function of activity $a$ for different values of $n$.
For decreasing $a$, we observe a clear transition from a fully turbulent to a completely  quiescent state.
The critical activity $a_n^\ast$ is determined numerically and marks the point where turbulent patches are finally unable to persist.
Here, $a_n^\ast$ is shifted to smaller activities for larger $n$, thus extending the region of hysteresis and coexistence.

When we plot $\Phi$ as a function of the distance to the critical point $(a - a_n^\ast)$, see the inset of Fig.~\ref{fig:transition}(b), we observe  power-law behavior $\Phi\propto (a - a_n^\ast)^{\beta^\ast}$.
We find that the exponent $\beta^\ast$ does not change when varying the steepness of the shear thinning crossover determined via the parameter $n$.
Fitting the curves in Fig.~\ref{fig:transition}, we obtain $\beta^\ast = 0.352(20)$, where the number in brackets denotes an error estimate corresponding to a confidence interval of $\SI{95}{\percent}$.
However, additional simulations, demonstrate that the exponent $\beta^\ast$ is in fact not universal, see Supplementary Note 5 and Supplementary Figures 3 and 4. 
Although independent or only weakly dependent on both $\nu_0/\nu_\infty$ and $n$, it depends on the value of the crossover shear rate $\zeta$.

The power-law scaling indicates that the transition may be linked to a universality class of non-equilibrium phase transitions.
Indeed, recent studies~\cite{takeuchi2007directed,sipos2011directed,lemoult2016directed,sano2016universal,doostmohammadi2017onset} have shown that emergence of turbulence in various systems can be understood as a transition of directed percolation (``DP'').
In this analogy, turbulent domains spreading to neighboring regions correspond to the excited state, whereas laminar domains correspond to the absorbing state.
Above a certain critical Reynolds number, the turbulent domains may persist in time, thus leading to a percolation transition. The direction of the spreading process of directed percolation therefore corresponds to the time dimension of the driven fluid flow.
In addition to driven Couette flows of passive systems~\cite{lemoult2016directed,chantry2017universal,klotz2022phase,gome2024phase}, directed percolation transitions have been found in channel flow~\cite{sano2016universal}, pipe flow~\cite{sipos2011directed}, turbulent liquid crystals~\cite{takeuchi2007directed}, as well as active nematics in microchannels~\cite{doostmohammadi2017onset}.

\begin{figure}
\includegraphics[width=0.9\linewidth]{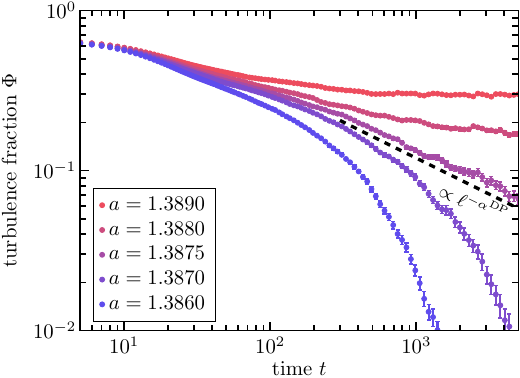}
\caption{\label{fig:quench}Critical-quench simulations. Evolution of the turbulence fraction $\Phi(t)$ after abruptly reducing the activity from $a=1.4$ to values close to the critical point. 
For activities larger than the critical value, $a > a^\ast$, $\Phi(t)$ saturates at a finite value, whereas for $a < a^\ast$, $\Phi(t)$ decays to zero.
At the critical activity, here $a^\ast \approx 1.3875$, we observe approximate power-law scaling, $\Phi(t) \propto t^{-\alpha}$, consistent with $2+1$ directed percolation.
This is indicated by the dashed black line, which represents the exponent of directed percolation $\alpha^{DP} = 0.4505(10)$~\cite{munoz1999avalanche,takeuchi2007directed}.
The remaining parameters are $\zeta=2$, $\nu_0 = 1.5 \nu_\infty$, and $n=4$. Error bars denote the standard error.}
\end{figure}

Motivated by these studies, we investigate whether the transition between a fully macroscopically quiescent system and patchy self-sustained turbulent states, as observed in our shear-thinning active fluid, exhibits features consistent with directed percolation as well.
As discussed, the power-law exponent $\beta^\ast$, governing the scaling of the turbulence fraction with $a-a^\ast_n$ depends on the magnitude of the crossover shear rate $\zeta$ of shear thinning.
Directly comparing exponents that govern the scaling in terms of the distance to the critical point with those of the directed percolation universality classes thus seems futile.
In fact, there is no a priori reason why the activity $a$ should linearly correspond to the spreading probability, which is the control parameter in directed percolation.
Thus, we rather focus our attention on the critical point directly and not on how it is approached.

In directed percolation, the spatial and temporal correlations become scale-free at the critical point~\cite{hinrichsen2000non}.
Accordingly and motivated by previous works~\cite{lemoult2016directed,takeuchi2007directed,chantry2017universal}, we thus focus on the distributions of certain characteristics of the absorbing state, which here is the quiescent state. Specifically, those characteristic parameters are the distances (gap lengths) between turbulent domains, $\ell$, and the time intervals (duration) of quiescence between the occurrences of turbulence at fixed spatial positions, $\tau$.
We employ the coarse-grained viscosity $\nu_\mathrm{cg}(\mathbf{x},t)$ to determine these quantities.
For example, the flow field exhibits a quiescent gap of length $\ell_x$ in $x$ direction if $\nu_\mathrm{cg}(x,y,t) < \nu_*(a)$ and $\nu_\mathrm{cg}(x+\ell_x,y,t) < \nu_*(a)$, but $\nu_\mathrm{cg}(x+\ell',y,t) > \nu_*(a)$ for $0 < \ell' < \ell_x$.
The gap length $\ell_y$ and quiescent time $\tau$ are determined analogously.

To obtain the distributions $\mathcal{P}(\ell)$ and $\mathcal{P}(\tau)$ at the critical point, we perform additional long-running, large-scale numerical evaluations for the sets of parameters explored so far.
Here, we choose values of the activity $a$ as close to the critical value $a_n^\ast$ as possible.
We first determine distributions of gap lengths for the $x$ and $y$ directions separately.
As the system exhibits isotropic symmetry in the statistical sense, see Supplementary Note 2, these distributions, $\mathcal{P}(\ell_x)$ and $\mathcal{P}(\ell_y)$, are equal for large enough sample sizes.
We thus calculate $\mathcal{P}(\ell)$ as the average of $\mathcal{P}(\ell_x)$ and $\mathcal{P}(\ell_y)$.
The quiescent time distribution, $\mathcal{P}(\tau)$, can be determined directly.

Figure~\ref{fig:DP} shows $\mathcal{P}(\ell)$ and $\mathcal{P}(\tau)$ for $\zeta = 2$, $\nu_0/\nu_\infty = 1.5$, different values of $n$, and activities as close to the critical point as possible.
Both distributions exhibit power-law scaling for sufficiently large $\ell$ and $\tau$, implying self-similarity at the critical point.
Here, the spatial and temporal critical exponents, $\mu_\perp$  and $\mu_\parallel$ , are defined via $\mathcal{P}(\ell)\propto\ell^{-\mu_\perp}$ and $\mathcal{P}(\tau)\propto\tau^{-\mu_\parallel}$.
Generally, for $2+1$ directed percolation (two spatial, one temporal dimension), the exponents are given by $\mu_\perp^\mathrm{DP} = 1.204(2)$ and $ \mu_\parallel^\mathrm{DP} = 1.5495(10)$~\cite{munoz1999avalanche,takeuchi2007directed}, as indicated by the dashed lines in Fig.~\ref{fig:DP}. 
The obtained distributions show convincing agreement with these exponents, indicating that the emergence of self-sustained, active turbulent states in shear-thinning fluids may indeed be linked to a directed percolation transition.
Additional simulations with different parameter values further support this conclusion, see Supplementary Note 5.

Exponents determined via fitting to the data points are close to the exponents expected for $2+1$ directed percolation as well. 
From corresponding fitting procedures for Fig.~\ref{fig:DP} and additional data points, see Supplementary Figure 4, we obtain values of the spatial exponent between $\mu_\perp = 1.172(5)$ and $\mu_\perp = 1.218(6)$ and of the temporal exponent between $\mu_\parallel = 1.5280(30)$ and $\mu_\parallel = 1.5860(32)$.
Further details on the obtained values and the fitting procedure are provided in the Methods section, including Table~\ref{tab:exponents}. 
We remark that the irregularities emerging in Fig.~\ref{fig:DP}(a) for the largest considered values of $\ell$ are due to the finite system size of $128\pi \times 128 \pi$.

As intuitively implied above, the link to a $2+1$ directed percolation transition is possibly given by the emergence and time evolution of the two-dimensional turbulent patches.
Numerical results show that these may spread or die out, suggesting a correspondence to the survival or decay of excited clusters in directed percolation.
To further establish this analogy, we perform additional ``critical-quench'' simulations to determine how turbulent patches decay.
For this purpose, we start at an intermediate activity of high turbulence fraction, abruptly decrease the activity to a value close to the critical point, and then observe the time evolution of the turbulence fraction $\Phi(t)$.
For every investigated magnitude of activity, these simulations are repeated $200$ times to obtain adequate statistics.
Figure~\ref{fig:quench} shows $\Phi(t)$ for different values of $a$ close to the critical point for an exemplary set of parameters. 
Below criticality, $\Phi(t)$ seems to decay exponentially to zero, whereas above criticality, $\Phi(t)$ saturates at a finite value.
At the critical point, we observe approximately algebraic decay according to a power law $\Phi(t) \propto t^{-\alpha}$ in the long-time limit.
In $2+1$ directed percolation, the corresponding exponent is given as $\alpha^\mathrm{DP} = 0.4505(10)$~\cite{munoz1999avalanche,takeuchi2007directed}.
Fitting $\alpha$ for $300 < t < 5000 $ in the situation presented in Fig.~\ref{fig:quench}, we obtain $\alpha = 0.4207(16)$. Thus, the decay of $\Phi(t)$ at criticality is roughly consistent with directed percolation.

Our results open the path for further exploration of the actual nature of the transition between quiescence and self-sustained patchy mesoscale turbulence.
We here find multiple features at the critical point that are consistent with directed percolation.
A major next step is to investigate how the activity $a$ relates to the control parameter in directed percolation.
In this context, key questions concern the spreading probability of turbulent patches as well as the motion of fronts separating turbulence and quiescence~\cite{gome2024phase}.\\

\begin{figure}
\includegraphics[width=0.95\linewidth]{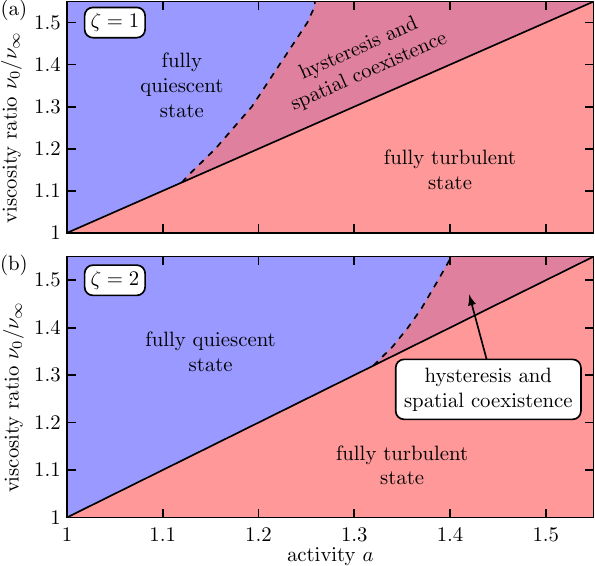}
\caption{\label{fig:stateDiagram}State diagrams. The observed states for the entire system as a function of activity $a$ and viscosity ratio $\nu_0/\nu_\infty$ for $n=4$ as well as (a) $\zeta = 1$ and (b) $\zeta = 2$. 
The quiescent state is linearly unstable for $a > \nu_0/\nu_\infty$ (diagonal line), resulting in a fully turbulent state. 
For activities below the linear instability, $a < \nu_0/\nu_\infty$, but right to the dashed line, there is a region of hysteresis and spatial coexistence between turbulent and quiescent regions.}
\end{figure}

\noindent \textbf{State diagrams}

\noindent Finally, to summarize, Fig.~\ref{fig:stateDiagram} shows state diagrams of the shear-thinning active suspension as a function of activity $a$ and viscosity ratio $\nu_0/\nu_\infty$ for different values of reference shear rate $\zeta$.
The linear instability of the quiescent state for $a > \nu_0/\nu_\infty$ yields the diagonal line, right of which we always encounter a fully turbulent state of the entire system.
Left of the diagonal, the hysteretic region of coexistence is found above a certain threshold activity.
Further to the left, we find the state of complete macroscopic quiescence. 
%Here, for these low activities, local shear rates do not reach the low-viscosity regime and coexistence between turbulence and quiescence is thus ruled out.

When comparing Figs.~\ref{fig:stateDiagram}(a) and (b), we observe that an increase in crossover shear rate
shifts the region of hysteresis and spatial coexistence to higher activity $a$.
We give an intuitive explanation in the following.
The turbulent and quiescent states correspond to low- and high-viscosity domains, respectively.
Both regimes must be accessible for the system, so that these two states can coexist.
However, for too low  activity, local shear rates mostly remain too low so that shear thinning does not set in effectively. Thus, the system does not reach the low-viscosity regime.
In particular, local shear rates must significantly exceed the crossover shear rate $\zeta$ for shear thinning to play a significant role.
Together with an increased value of $\zeta$, the minimum required activity to facilitate self-sustained turbulence based on shear thinning thus becomes larger. Consequently, the coexistence region in the state diagram is shifted to the top right.\\

\noindent {\large \textbf{Discussion}}\\[-0.5\baselineskip]

\noindent Summarizing, we reveal heterogeneous spatial coexistence of self-sustained turbulent and macroscopically quiescent regions in shear-thinning active suspensions. These states are found in emergent hysteretic regimes of mesoscale turbulence as a function of activity. Concerning the associated anomalous velocity statistics, we mention related experimental observations on bacterial suspensions of \textit{Bacillus subtilis}~\cite{benisty2015antibiotic,ilkanaiv2017effect,be2020phase}. There,  motile and immotile areas coexist, induced, for example, by the presence of sublethal doses of antibiotics~\cite{benisty2015antibiotic} or elevated aspect ratios~\cite{be2020phase}. We here provide an illustrative explanation via non-Newtonian shear thinning.  
For instance, excretions of bacteria \cite{hall2004bacterial,worlitzer2022biophysical,jana2020nonlinear} can induce such shear thinning in biological suspensions. Our explanation is not based on density variations as involved, for example, in motility-induced phase separation~\cite{worlitzer2021motility,worlitzer2021turbulence}.

Concerning experimental realizations, it is useful to classify the explored range of activities $a$.
In a recent study~\cite{slomka2017geometry}, the turbulent states obtained by solving the equation for a Newtonian suspension are compared with the spatiotemporal patterns in bacterial suspensions of $\textit{Bacillus subtilis}$. For example, the parameter choice $\Gamma_0 = \SI{e3}{\square\micro\meter\per\second}$, $\Gamma_2/\Gamma_0 = \SI{1.24e2}{\square\micro\meter}$ and $\Gamma_4/\Gamma_0 = \SI{3.53e2}{\raiseto{4}\micro\meter}$ leads to good agreement. In our rescaled equations, this choice corresponds to a dimensionless activity of $a \approx 1.1$.
This value is in the range of the activities investigated in our study.
Since we have observed the same qualitative behavior for all parameter sets, we expect experimental studies aiming to investigate the impact of shear thinning on active suspensions to be feasible.
Our work may serve as a guide to choose shear-thinning solvent media with appropriate properties.

Obviously, Eqs.~(\ref{eq:NSt}) and (\ref{eq:stressExpansion}) reproduce well the phenomenology of active suspensions as observed in experiments. This was confirmed quantitatively in a recent study~\cite{slomka2017spontaneous}, where reasons for this match were discussed in detail.
In particular, the theory including higher-order Laplaceans in the stress tensor captures the selection of a characteristic vortex size while the presence of the nonlinear advection term leads to turbulence-like dynamics.
One may raise the question about the significance of the nonlinear term in the context of microswimmer suspensions, which contain microscopic objects of micrometer size. 
In favor of its relevance, the effective viscosity of the overall suspension on length scales larger than the microswimmers can be substantially lowered in active suspensions~\cite{haines2008effective,sokolov2009reduction,heidenreich2016hydrodynamic,lopez2015turning}.
Moreover, the characteristic length scale of collective motion, that is, the vortex size, is much larger than single microswimmers~\cite{sokolov2012physical,wensink2012meso,nishiguchi2018engineering}. Also the speed of collective motion can be larger than the speed of individual microswimmers~\cite{nishiguchi2018engineering}. 
All these effects, a lower overall viscosity, larger relevant length scales, as well as increased collective speed of motion, contribute to an increase in the effective Reynolds number. 
We refer to Supplementary Note 1 for a more detailed discussion.
Moreover, we note that the nonlinear term is not only connected to inertial effects in the classical sense. Particularly, in the classical sense, it is associated with advective transport. Therefore, in the present context, it also expresses the fact of active transport induced by the active agents. This active transport can certainly play a major role in thin films of active suspensions. There is a previous consideration of this context \cite{linkmann2020condensate}. Corresponding active contributions stress the importance of the nonlinear term and imply an additional increase in the effective Reynolds number.

The description may find its application also in further contexts, beyond the field of active suspensions. 
For instance, expansions of the stress tensor of the kind shown in Eq.~(\ref{eq:stressExpansion}) may provide useful approximative characterizations of emerging patterns in various other types of fluids, for example, in the case of magnetically driven flow~\cite{ouellette2007curvature} or, more generally, forced~\cite{jimenez2007spontaneous} and instability-driven turbulence~\cite{van2024vortex}.
Furthermore, the one-dimensional version of Eqs.~(\ref{eq:NSt}) and (\ref{eq:stressExpansion}), denoted as the Nikolaevskiy model~\cite{beresnev1993model}, arises in the context of seismic waves~\cite{beresnev1993model,tribelsky1996new} and reaction-diffusion systems~\cite{tanaka2004chemical}.

From a fundamental perspective, our study extends recent attempts of linking nonequilibrium transitions, such as the emergence of turbulence~\cite{sipos2011directed,lemoult2016directed,doostmohammadi2017onset,reinken2024pattern}, to universality classes of statistical physics, ranging from Ising behavior~\cite{marcq1997universality,reinken2022ising} to directed percolation~\cite{takeuchi2007directed,sipos2011directed,lemoult2016directed,sano2016universal,doostmohammadi2017onset}.
In the more practical context of applications, the observed phenomena are
relevant for microfluidic and mesoscale mixing based on active suspensions~\cite{kim2004enhanced,sokolov2009enhanced,leptos2009dynamics,sanjay2020friction,mukherjee2021anomalous,reinken2022optimal}. Mesoscale turbulence enhances mixing. We have demonstrated that, through shear thinning, the regime of mesoscale turbulence can be extended to lower activities, as the system becomes heterogeneous and maintains self-sustained turbulent patches. 
In a way, the situation reminds of type-II superconductors, where, in analogy, applications can be extended to higher magnetic fields, maintaining superconductivity through emergent spatial heterogeneities in the system~\cite{degennes2018superconductivity}.\\[1\baselineskip]

\noindent {\large \textbf{Methods}}\\[-0.5\baselineskip]

\noindent \textbf{Linear stability analysis}

\noindent The linear stability analysis for the rescaled Navier--Stokes equation, that is, Eq.~(\ref{eq:generalizedNStRescaled}), is provided in the following.
In particular, we consider the stability of the macroscopically quiescent solution $\mathbf{v}_0 = \mathbf{0}$.

As a first step, we add small perturbations to velocity and pressure,
\begin{equation}
\label{eq:solutionPlusPerturbation}
\mathbf{v}(\mathbf{x},t) = \mathbf{v}_0 + \delta \mathbf{v}(\mathbf{x},t)\, , \quad \tilde{p}(\mathbf{x},t) = \tilde{p}_0 + \delta \tilde{p}(\mathbf{x},t)\, .
\end{equation}
To continue, we make the ansatz
\begin{equation}
\label{eq:perturbation}
\delta \mathbf{v}(\mathbf{x},t) = \delta \hat{\mathbf{v}} \, e^{\lambda t + i \mathbf{k}\cdot \mathbf{x}}\, , \quad \delta \tilde{p}(\mathbf{x},t) =  \delta \hat{p} \, e^{\lambda t + i \mathbf{k}\cdot \mathbf{x}}\, ,
\end{equation}
where $\lambda$ is the complex growth rate and $\mathbf{k}$ is the wavevector of the perturbation.
We insert Eqs.~(\ref{eq:solutionPlusPerturbation}) and (\ref{eq:perturbation}) into Eq.~(\ref{eq:generalizedNStRescaled}) %of the main text 
and linearize around the quiescent solution.
Evaluating the temporal and spatial derivatives, we obtain
\begin{equation}
\label{eq:LSAStep1}
\lambda \delta \hat{\mathbf{v}} = - i \mathbf{k} \delta \hat{p} - \left(\nu_0 |\mathbf{k}|^2 /\nu_\infty - 2 a |\mathbf{k}|^4 + a |\mathbf{k}|^6\right) \delta \hat{\mathbf{v}}  \, .
\end{equation}
Multiplying by $\mathbf{k}$ and using the incompressibility condition $\mathbf{k} \cdot \delta \hat{\mathbf{v}} = 0$ implies that $\delta \hat{p} = 0$.
Thus, the growth rate $\lambda$ as a function of the wavevector is determined as
\begin{equation}
\label{eq:growthRate}
\lambda(\mathbf{k}) = - \nu_0 |\mathbf{k}|^2 /\nu_\infty + 2 a |\mathbf{k}|^4 - a |\mathbf{k}|^6\, .
\end{equation}
We note that $\lambda(\mathbf{k})$ is always real 
and find that 
it can become positive for 
\begin{equation}\label{eq:ath}
a > a_\mathrm{th}=\nu_0/\nu_\infty.
\end{equation}
This defines the threshold value $a_\mathrm{th}$  
as introduced in the main text. At this value, a finite-wavelength instability sets in and modes associated with the wavenumber $k_\mathrm{c} = 1$ start to grow.
Increasing $a$ above $a_\mathrm{th}$, a band of unstable modes develops. The wavenumber of the mode of maximum growth rate $k_\mathrm{m}$ is close to the critical wavenumber $k_\mathrm{c}$.\\

\noindent \textbf{Numerical methods}

\noindent We employ a pseudo-spectral scheme to solve Eq.~(\ref{eq:generalizedNStRescaled}) %of the main text 
in a two-dimensional system with periodic boundary conditions.
Here, gradient terms are computed in Fourier space, which significantly speeds up the calculations~\cite{canuto2007spectral}.
Time integration is performed via a fourth-order Runge-Kutta method with a time step of $\Delta t = 0.05$. 
It is combined with an operator splitting technique treating the linear and nonlinear parts consecutively~\cite{cross2009pattern}. 
This approach involves the following process for every Runge-Kutta step.
We first consider only the nonlinear terms in the evolution equation, which we integrate to obtain an intermediate velocity field.
This field is then used as the initial condition for a second evolution problem using only the remaining linear terms.
In this Runge--Kutta step, we first ignore the pressure $\tilde{p}$ .
Instead, $\tilde{p}$ is subsequently obtained via a projection method to ensure that the incompressibility condition, $\nabla \cdot \mathbf{v} = 0$ is fulfilled~\cite{canuto2007spectral}.
Here, we compute the divergence of the velocity field obtained via the Runge--Kutta step as describe above.
The pressure is then determined in Fourier space as the solution of a Poisson equation that renders the velocity field divergence-free~\cite{canuto2007spectral}.

The main results
are obtained by starting the numerical calculations for an activity $a$ that is large enough for turbulence to develop. That is, we select $a > \nu_0/\nu_\infty$, implying that the macroscopically quiescent state is linearly unstable.
The initial conditions are set to  $\mathbf{v}(\mathbf{x},t) = \mathbf{0}$, with small random perturbations added to each velocity component at every grid point, following a uniform distribution over $[-0.01,0.01]$.
We then decrease the activity $a$ in small steps, let the calculations run for $500$ time units, and store the resulting velocity fields for subsequent calculations.
These fields are used as initializations for further long-time calculations, which produce the presented results within the hysteretic regime.
We have confirmed that the results do not depend on the specific numerical procedure.
For example, starting directly from an activity $a$ within the hysteretic regime and adding a sufficiently strong local perturbation (such as a localized lattice-like arrangement of vortices) results in patchy turbulent states as well, see Supplementary Note 6.

Before analyzing the dynamics, we always let the simulations for a set of parameters run for at least $3000$ time units to ensure the development of statistically stationary states.
We have confirmed that these long waiting times are sufficient to exclude transient behavior, see Supplementary Note 6.
Close to the critical point, simulations are run even longer (up to $30000$ time units).
The mean enstrophy $\langle \omega^2\rangle$, the velocity distribution $\mathcal{P}(v)$, and the turbulence fraction $\Phi$ are averaged over $5$ consecutive time intervals of $1500$ time units. 
Error bars in the figures denote the standard error, that is, the standard deviation divided by the square root of the number of time intervals.

The velocity statistics $\mathcal{P}(v)$ as shown in Fig.~\ref{fig:snapshots}(d) are determined as an average over $\mathcal{P}(v_x)$ and $\mathcal{P}(v_y)$, that is, the distributions for $x$ and $y$ velocity components, respectively. 
Due to the statistical dynamical isotropy of the turbulent state, see Supplementary Note 2, the two distributions are the same within our statistical errors.
The same holds for the quiescent gap length distribution $\mathcal{P}(\ell)$, which is determined as an average over $\mathcal{P}(\ell_x)$ and  $\mathcal{P}(\ell_y)$.
The distributions $\mathcal{P}(\tau)$ and $\mathcal{P}(\ell)$ are determined as close as possible to the critical point to investigate the transition regarding a possible link to the directed percolation universality class.
Here, we analyze the dynamics in a window of at least $10000$ time units.

For most calculations, the system size is set to $128\pi\times 128\pi$ and the spatial resolution to $768\times 768$ grid points.
We have checked that the systems considered are large enough so that finite-size effects do not play any obvious role, see Supplementary Note 7 and Supplementary Figure 6.
To obtain the state diagrams shown in Fig.~\ref{fig:stateDiagram}, we rather use a system size of $64\pi\times 64\pi$ and $384\times 384$ grid points.
These sizes are quite large compared to the vortex size, which is close to the critical length scale $2 \pi/k_\mathrm{c} = 2 \pi$.
To ensure stability, we only explore activity regimes of $a<1.55$~\cite{slomka2015generalized}.\\

\noindent \textbf{Coarse-graining prodecure}

\noindent To analyze the coexistence of turbulent and macroscopically quiescent regions in space, we first employ a coarse-graining procedure to smoothen the locally nonuniform vortex patterns.
In particular, we average the local viscosity over a square region of side length $\Delta x$ around each point in space.
The resulting coarse-grained viscosity field $\nu_\mathrm{cg}(\mathbf{x},t)$ at location $\mathbf{x}=(x,y)$ is thus obtained from the field $\nu(\mathbf{x},t)$ via
\begin{equation}
\nu_\mathrm{cg}(\mathbf{x},t) = \frac{1}{\Delta x^2} \int_{y-\Delta x/2}^{y+\Delta x/2} \ \int_{x-\Delta x/2}^{x+\Delta x/2} \nu(\tilde{\mathbf{x}},t) \, \mathrm{d}\tilde{x}\, \mathrm{d}\tilde{y} \, ,  
\label{eq:coarse-graining}
\end{equation}
where $\tilde{\mathbf{x}} = (\tilde{x},\tilde{y})$.
As the vortex structures emerge from the finite-wavelength instability at $a=\nu_0/\nu_\infty$, we here choose a side length $\Delta x$ equal to the critical length scale, that is, $\Delta x = 2\pi/k_\mathrm{c} = 2\pi$.
The more refined structure of patterns is thus averaged out.

When investigating the spatial coexistence of turbulent and macroscopically quiescent domains, we apply the relations derived above for global linear stability 
to the local scale.
Locally, the quantity of interest is the coarse-grained viscosity $\nu_\mathrm{cg}(\mathbf{x},t)$.
%We rearrange the stability condition for the quiescent state, 
For our purpose, we replace in Eq.~\eqref{eq:ath} $\nu_0$ by $\nu_\mathrm{cg}(\mathbf{x},t)$. Then, approximately, we expect a quiescent domain to be stable if, for that domain, $\nu_\mathrm{cg}(\mathbf{x},t)>\nu^\ast(a) = a \nu_\infty$.
Contrarily, if $\nu_\mathrm{cg}(\mathbf{x},t) < \nu^\ast= a \nu_\infty$, local perturbations grow and turbulence can sustain itself in this area. Overall, this condition allows to distinguish between turbulent and quiescent regions. It forms the basis of the color scale that we have chosen for illustration in Fig.~2(e) and (f). % of the main text.
Green color indicates quiescent domains of $\nu_\mathrm{cg}(\mathbf{x},t) > \nu^\ast$, whereas purple color marks turbulent regions of $\nu_\mathrm{cg}(\mathbf{x},t) < \nu^\ast$.\\[1\baselineskip]

\begin{table*}[t]
\caption{\label{tab:exponents}Critical exponents. The exponents $\mu_\perp$ and $\mu_\parallel$ of the power-law scalings $\mathcal{P}(\ell)\propto\ell^{-\mu_\perp}$ and $\mathcal{P}(\tau)\propto\tau^{-\mu_\parallel}$ for different sets of parameter values very close to the critical point. $\mathcal{P}(\ell)$ and $\mathcal{P}(\tau)$ are the distributions of the spatial distance between turbulent patches (gap lengths) $\ell$  and of the local quiescent time intervals between the occurrences of turbulence $\tau$, respectively. Values for the exponents are obtained via fitting the scaling laws of directed percolation (DP) to the data points in the intervals listed in the table.
The exponent $\mu_\perp$ refers to the spatial dimension and $\mu_\parallel$ to the direction of percolation, that is, the temporal dimension.
For comparison, the exponents $\mu_\perp^\mathrm{DP}$ and $\mu_\parallel^\mathrm{DP}$ for $2+1$ directed percolation~\cite{takeuchi2007directed,hinrichsen2000non} are included as well.
Error estimates correspond to $\SI{95}{\percent}$ confidence intervals.\\}
\renewcommand{\arraystretch}{1.5} 
\begin{tabular}{|c|c|c|c|l|l|r|r|}
\hline
$ \quad \ \zeta \quad \ $ & $\ \ \nu_0/\nu_\infty \ \ $ & $ \quad n \quad $ & $ \qquad \ a \qquad \ $ & $ \qquad \ \mu_\perp \qquad \ $ & $ \qquad \quad \mu_\parallel \qquad \quad $ & \multicolumn{2}{c|}{regions used for parameter fit} \\ \hline \hline
$1$ & $1.3$ & $4$ & $1.19399$ & \quad $1.197(19)$ & $ \quad 1.5860(32)$ &  \ $ 50 < \ell < 200 $ \ & \ $ 30 < \tau < 500 $ \ \\  \hline
$1$ & $1.4$ & $4$ & $1.22216$ & \quad $1.218(6)$ & \quad $ 1.5280(33)$ &  $ 30 < \ell < 200 $ \ &  $ 30 < \tau < 500 $ \ \\  \hline
$1$ & $1.5$ & $4$ & $1.24417$ & \quad $1.172(5)$ & \quad $ 1.5758(54)$ &  $ 20 < \ell < 200 $ \ &  $ 30 < \tau < 500 $ \ \\  \hline
$2$ & $1.5$ & $2$ & $1.41460$ & \quad $1.200(6) $ & \quad $1.5424(96)$  & \ $50 < \ell < 300$ \ &  $90 < \tau <400$  \ \\  \hline
$2$ & $1.5$ & $4$ & $1.38825$ & \quad $1.196(4) $  &  \quad $1.5596(76)$  & \ $50 < \ell < 150$ \  &  $50 < \tau < 300$ \ \\  \hline
$2$ & $1.5$ & $8$ & $1.36585$ & \quad $1.211(8) $ & \quad $1.5443(51)$ & \, $110 < \ell < 300$ \ &  $50 < \tau < 700$ \ \\ \hline \noalign{\vskip\doublerulesep\vskip-\arrayrulewidth} \cline{1-6}
\multicolumn{4}{|c|}{\ $\mu_\perp^\mathrm{DP}$ and $\mu_\parallel^\mathrm{DP}$ for $2+1$ directed percolation \ } & \quad $1.204(2)$ & \quad $1.5495(10)$ & \multicolumn{2}{c}{} \\ \cline{1-6}
\end{tabular}
\end{table*}

\noindent \textbf{Fitting critical exponents}

\noindent In a directed percolation transition, the spatial and temporal structure of the system becomes scale-free at the critical point.
This is observed in the distributions $\mathcal{P}(\ell)$ and $\mathcal{P}(\tau)$ of the size $\ell$ and duration $\tau$ of gaps between excited (here corresponding to turbulent) domains.
At the critical point, these distributions follow power laws, $\mathcal{P}(\ell)\propto\ell^{-\mu_\perp}$ and $\mathcal{P}(\tau)\propto\tau^{-\mu_\parallel}$. 
In directed percolation, the critical exponents $\mu_\perp$ and $\mu_\parallel$ are related to the exponents of perpendicular (spatial) and parallel (temporal) correlation lengths, $\nu_\perp^\mathrm{DP}$ and $\nu_\parallel^\mathrm{DP}$, respectively, via $\mu_\perp^\mathrm{DP} = 2 - \beta^\mathrm{DP}/\nu_\perp^\mathrm{DP}$ and $\mu_\parallel^\mathrm{DP} = 2 - \beta^\mathrm{DP}/\nu_\parallel^\mathrm{DP}$~\cite{takeuchi2007directed,lemoult2016directed}.
There, $\beta^\mathrm{DP}$ is the critical exponent of the proportion of excited regions~\cite{takeuchi2007directed,lemoult2016directed}.

So far, a direct linear correspondence between the control parameter in directed percolation and the activity or other parameters in the type of mesoscale turbulence addressed above has not been revealed.
When comparing the results, we thus focus on the behavior at the critical point directly, which in directed percolation is determined by the two exponents $\mu_\perp^\mathrm{DP}$ and $\mu_\parallel^\mathrm{DP}$.
To obtain an accurate estimate of
corresponding exponents in our active, shear-thinning system of mesoscale turbulence, we move as close to the critical point as possible.
Fitting the curves $\mathcal{P}(\ell)$ and $\mathcal{P}(\tau)$ shown in Fig.~\ref{fig:DP} and Supplementary Figure 4, we obtain the critical exponents for the six parameter sets on display.
The results are summarized in Table~\ref{tab:exponents}, which includes the ranges of $\ell$ and $\tau$ used for the fit.
We find that the exponents obtained from the fits to our data are consistent with the expected values for $2+1$ directed percolation.\\

\noindent \textbf{Data availability}\\
\noindent The data in support of the reported findings are available within the paper and its supplementary information files.\\

\noindent \textbf{Code availability}\\
\noindent The computer codes used for the numerical calculations and analysis are published on the repository Zenodo and can be found at \url{https://doi.org/10.5281/zenodo.15718824}.\\

%\bibliography{references}


\vspace{1\baselineskip}

\hspace{1.cm}

\noindent \textbf{Acknowledgements}\\
\noindent The authors thank  S\'ebastien Gom\'e for stimulating discussions and the Deutsche Forschungsgemeinschaft (German Research Foundation, DFG) for support through the Research Grant no.\ ME 3571/5-1 (project number 413993436). Moreover, A.M.M. acknowledges support by the DFG through the Heisenberg Grant no.\ ME 3571/4-1 (413993216).  \\

\noindent This version of the article has been accepted for publication, after peer review but is not the Version of Record and does not reflect post-acceptance improvements, or any corrections. The Version of Record is available online at: \url{http://dx.doi.org/10.1038/s42005-025-02200-3}.\\

\noindent \textbf{Author contributions}\\
\noindent A.M.M.\ and H.R.\ designed the study, wrote the paper, and discussed the results. H.R.\ developed and performed the computational calculations. \\

\noindent \textbf{Competing interests}\\
\noindent The authors declare no competing interests.

\end{document}

\title{Self-sustained patchy turbulence in shear-thinning active fluids\\[0.5\baselineskip] \textit{Supplementary Information}}

\author{Henning Reinken}
\email{henning.reinken@ovgu.de}
\affiliation{Institut f\"ur Physik, Otto-von-Guericke-Universität Magdeburg, Universitätsplatz 2, 39106 Magdeburg, Germany} 

\author{Andreas M. Menzel}
\email{a.menzel@ovgu.de}
\affiliation{Institut f\"ur Physik, Otto-von-Guericke-Universität Magdeburg, Universitätsplatz 2, 39106 Magdeburg, Germany}

\date{\today}

\begin{abstract}
In this Supplementary Information, we provide some additional background information on the theoretical description, as well as supplementary data that further supports the findings already discussed in the main text.
In particular, we show that the patchy turbulent state exhibits statistical isotropy.
We discuss the distribution function of the coarse-grained viscosity, which illustrates the coexistence of turbulent and quiescent regions.
Further, we provide a description of the Supplementary Movies.
Finally, we present results for different parameter values and different system sizes, which do not provide additional insight but show the robustness of our results.
\end{abstract}

\maketitle

\section{Classification of the theoretical description}

The theoretical framework we employ was introduced in a recent study~\cite{slomka2015generalized} to characterize states of mesoscale turbulence in suspensions of active microswimmers.
It is formulated in terms of the overall velocity field of the suspension.
As such, it represents a modified version of the Navier--Stokes equation.
The effects of the active microswimmers enter via the stress tensor. 
Equation.~(2) in the main text represents the simplest isotropic formulation that selects patterns of a characteristic length scale and implies stability at small and large wavenumbers~\cite{slomka2017geometry,slomka2017spontaneous}.

For studies on individual microswimmer dynamics, the Stokes flow approximation is often employed, which neglects inertial effects.
This approximation is usually motivated by the Reynolds number calculated from the typical length scale and propulsion speed of a single microswimmer.
However, as discussed in a recent article~\cite{slomka2017spontaneous}, the low-Reynolds-number assumption is violated when considering the dynamics of the whole suspension.
We thus retain the nonlinear advection term in the Navier--Stokes equation.
To illustrate this, we consider the example of bacterial suspensions.
Here, the collective velocity can exceed the speed of a single bacterium by up to an order of magnitude~\cite{sokolov2007concentration} and the vortex size is much larger than individual bacteria~\cite{sokolov2012physical,wensink2012meso}, for example, about $10$ times as large in \textit{Bacillus Subtilis}~\cite{nishiguchi2018engineering}. 
In addition, the effective viscosity may be significantly reduced due to activity~\cite{sokolov2009reduction}.
This effect follows from induced reorientations of microswimmers in flow gradients.
The resulting active flows strengthen the shear flow in suspensions of pusher-like swimmers (including most bacteria).
To lowest order, this effect is linear in the flow gradients~\cite{heidenreich2016hydrodynamic}, similar to viscous dissipation in the Newtonian case, but with opposite sign.
In combination, they lead to a constant reduction in effective viscosity, to this order independent of the shear rate~\cite{heidenreich2016hydrodynamic}.
Such a reduction in viscosity can be more than an order of magnitude in size~\cite{sokolov2007concentration}, even proposing apparently inviscid bacterial superfluids~\cite{lopez2015turning}.
Onto this effect of reduced viscosity, we here impose additional shear thinning due to the carrier fluid. By construction, shear thinning depends on the shear rate. 
In combination, these phenomena result in an effective Reynolds number that cannot be considered as small any longer. 
For example, considering a typical vortex size of $\sim \SI{100}{\micro\meter}~$~\cite{nishiguchi2018engineering}, a collective speed of $\sim \SI{100}{\micro\meter\per\second}~$~\cite{nishiguchi2018engineering}, and a reduced viscosity of $\sim \SI{10000}{\micro\meter\squared\per\second}$ ($\sim 1\%$ of the viscosity of water), we reach a Reynolds number of $\mathrm{Re} \sim 1$.
In this case, inertial effects cannot be neglected and maintaining the term $\mathbf{v}\cdot \nabla \mathbf{v}$ seems justified and necessary.

Apart from that, the active particles themselves, when acting collectively, can induce local active transport of the overall suspension. Corresponding transport is quantified by an advective operator $\mathbf{v}\cdot\nabla$. It adds to the term $\mathbf{v}\cdot\nabla\mathbf{v}$ in the flow equation. In passive hydrodynamics it reflects inertial effects. Similar arguments associating the term $\mathbf{v}\cdot\nabla\mathbf{v}$ with active transport were listed in a different work~\cite{linkmann2020condensate}, where the authors included a corresponding calculation to support the theoretical description~\cite{slomka2015generalized, slomka2017geometry, slomka2017spontaneous}. At least for pusher-type microswimmers, an increased additional prefactor for the term $\mathbf{v}\cdot\nabla\mathbf{v}$ was proposed. This is in line with the reduction in viscosity mentioned above, which also refers to pusher-type microswimmers. Rescaling time, the increased prefactor of the term $\mathbf{v}\cdot\nabla\mathbf{v}$ can be recast to unity, so that we recover Eq.~(1) in the main text. Yet, this step is accompanied by another increase in effective, overall Reynolds number.

Our study assumes incompressibility of the suspension, $\nabla \cdot \mathbf{v} = 0$.
This is justified on the basis of typical solvents such as water, which are approximately incompressible. 
Since biological microswimmers themselves consist of water to a large extent, the overall velocity field (solvent plus microswimmers) can be assumed to be divergence-free as well.

We would also like to point out the differences between the employed framework and related theories of active systems.
We start with the Toner--Tu equations~\cite{toner2005hydrodynamics}, a continuum analog of the Vicsek model~\cite{vicsek1995novel}.
The Toner--Tu equations are formulated in terms of the velocity field that describes swarming states of active objects in dry active systems.
In other words, the presence of a solvent medium is not considered.
As such, more terms enter when compared to our theoretical description, which ensures both momentum conservation and Galilean invariance.
Recently, various studies used a framework denoted as Toner--Tu--Swift--Hohenberg equations to address mesoscale turbulence~\cite{wensink2012meso,reinken2024pattern,alert2021active,mukherjee2023intermittency}.
These descriptions are formulated in terms of the velocity field of the microswimmers as well.
Importantly, both the Toner--Tu and Toner--Tu--Swift--Hohenberg equations include a nonlinear advection term, which is crucial for the development of mesoscale turbulence in these descriptions.

In the present study, we aim to investigate the impact of complex rheological conditions on the mesoscale-turbulent state.
Compared to the theories mentioned above, the employed framework is particularly suitable in this context.
Since it is formulated in terms of the velocity field of the overall suspension and not in terms of the velocity field of only the microswimmers, the prefactor of the lowest-order term in the modified stress tensor directly corresponds to the overall viscosity.
Combining this description with non-Newtonian fluid models, such as the shear-thinning Cross fluid in our case, is thus straightforward.

\section{Statistical isotropy of the patchy turbulent state}

\begin{figure}[h]
\includegraphics[width=0.6\linewidth]{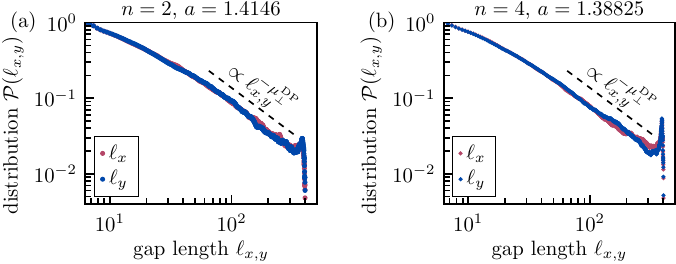}
\caption{\label{fig:statisticalIsotropy}Statistical isotropy of the patchy turbulent state concerning its scaling behavior.
Very close to the critical point, we show distributions $\mathcal{P}$ of  the distances (lengths of the gaps) between the turbulent regions in $x$ and $y$ direction, $\ell_x$ and $\ell_y$. 
Both distributions are normalized by the value at smallest $\ell$.
We show two exemplary plots: (a) $n=2$, $a=1.41460$ and (b) $n=4$, $a=1.38825$. 
The remaining parameters are $\zeta = 2$ and $\nu_0 = 1.5 \nu_\infty$. 
Dashed lines represent the exponent of $2+1$ directed percolation, $\mu_\perp^\mathrm{DP} = 1.204(2)$~\cite{takeuchi2007directed,hinrichsen2000non}.}
\end{figure}

In the main text, we discuss the features of the spatial and temporal structures very close to the critical point in terms of the distributions $\mathcal{P}$ of the distances (lengths of the gaps) between the turbulent regions $\ell$ and the time intervals of local quiescence $\tau$.
In this context, we show a single distribution $\mathcal{P}(\ell)$ instead of separate plots for the gap distances $\ell_x$ and $\ell_y$ in the two spatial directions.
The distribution $\mathcal{P}(\ell)$ is obtained by averaging over the distributions $\mathcal{P}(\ell_x)$ and $\mathcal{P}(\ell_y)$.
This approach is justified for statistical isotropy.
In the following, we show that this condition is satisfied in the patchy turbulent state discussed in our study.
To this end, Supplementary Figure~\ref{fig:statisticalIsotropy} includes the separate distributions $\mathcal{P}(\ell_x)$ and $\mathcal{P}(\ell_y)$ for two exemplary sets of parameters.
We observe the same power-law scaling behavior for both directions, implying statistical isotropy of the patchy turbulent state in view of the scaling behavior.

\section{Distribution function of the coarse-grained viscosity}
\label{app:coarseGraining}

Snapshots of the coarse-grained viscosity $\nu_\mathrm{cg}(\mathbf{x},t)$ are included in Fig.~2 of the main text. As discussed there, they clearly show the presence of both low- and high-viscosity regions, corresponding to the turbulent and macroscopically quiescent state, respectively.
To further illustrate our observations, we additionally determine the distribution $\mathcal{P}(\nu_\mathrm{cg}/\nu_\infty)$.  
It quantifies how much of the system exhibits high or low viscosity.
This distribution is plotted in Supplementary Figure~\ref{fig:viscosityDistribution} for two values of activity, $a=1.5$ and $a=1.41$. 
For the larger activity $a=1.5$, there is only one maximum in $\mathcal{P}(\nu_\mathrm{cg}/\nu_\infty)$. It reflects a fully turbulent system.
However, for the smaller activity $a=1.41$, we observe that a second maximum has emerged, here at $\nu_\mathrm{cg}/\nu_\infty = 1.5$. It corresponds to the quiescent state.
The presence of two maxima implies that there are separate, spatially coexisting regions of turbulence and of quiescence.

\begin{SCfigure}[0.8]
\includegraphics[width=0.42\linewidth]{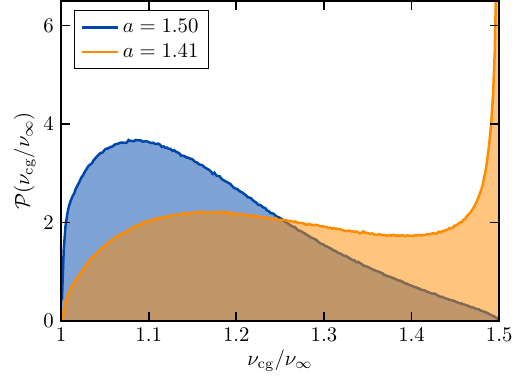}
\caption{\label{fig:viscosityDistribution}Distribution function $\mathcal{P}(\nu_\mathrm{cg}/\nu_\infty)$ for the coarse-grained viscosity $\nu_\mathrm{cg}$ for two different example values of activity. For larger values, here $a=1.5$, we observe a single maximum. It reflects a fully turbulent state. For lower values, here $a=1.41$, two maxima are present, They point to spatially coexisting regions of low and high viscosity, corresponding to turbulence and to quiescence, respectively.
The other parameters are set to $\nu_0/\nu_\infty = 1.5$, $n=4$, and $\zeta = 2$.}
\end{SCfigure}

\section{Description of Supplementary Movies}

Supplementary Movie 1 shows the dynamics of the vorticity field $\omega(\mathbf{x},t) = [\nabla \times \mathbf{v}(\mathbf{x},t)]_z$ in the statistically stationary state at the parameter values of $a = 1.39$, $\nu_0/\nu_\infty = 1.5$, $n=4$, and $\zeta = 2$. 
The size of the displayed area is $64\pi \times 64\pi$, corresponding to a quarter of the simulated system.
On the right-hand side, the plot is complemented by the corresponding coarse-grained viscosity $\nu_\mathrm{cg}(\mathbf{x},t)$ at the same time steps.
The movie illustrates the coexistence of turbulent and macroscopically quiescent regions in the hysteretic regime and how these domains rearrange over time.
In contrast to that, Supplementary Movie 2 shows how the fields $\omega(\mathbf{x},t)$ and $\nu_\mathrm{cg}(\mathbf{x},t)$ develop when the activity is suddenly increased to $a = 1.46$ when starting from the dynamics at $a = 1.39$.
Here, the turbulent domains grow until the quiescent regions vanish and the whole system exhibits turbulence.

\section{Additional data with different parameter values}

In Figs.~1, 2, and 3 of the main text, we focus on the presentation of data for a limited set of parameter values, namely a crossover shear rate of $\zeta = 2$, a viscosity ratio of $\nu_0/\nu_\infty=1.5$, and a steepness of the crossover $n=2$, $4$, and $8$.
Yet, as indicated, the results are rather general. 
To illustrate this point, we here include additional data when varying the parameter values.

\begin{figure}
\includegraphics[width=0.6\linewidth]{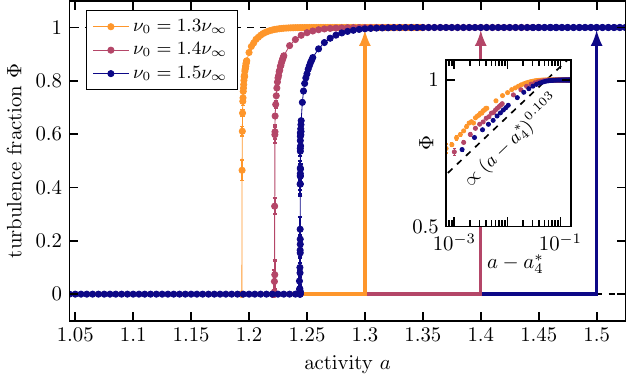}
\caption{\label{fig:transitionSuppl}Time-averaged area fraction of turbulent domains $\Phi$ as a function of activity $a$ for the crossover shear rate $\zeta = 1$ of shear thinning, degree of sharpness of the crossover $n=4$, and different values of viscosity ratio $\nu_0/\nu_\infty$. 
The hysteresis loops are indicated by the arrows with the same color as the associated data points.
Both the activity for which the global quiescent state becomes unstable, $a_\mathrm{th}$, and the critical activity, $a_{4}^\ast$, depend on $\nu_0/\nu_\infty$.
Inset: $\Phi$ as function of the distance to $a-a_4^\ast$, showing power-law scaling, that is, $\Phi \propto (a-a_4^\ast)^{\beta^\ast}$.}
\end{figure}

Supplementary Figure~\ref{fig:transitionSuppl} shows the turbulence fraction $\Phi$ as a function of activity $a$ for $\zeta = 1$, $n=4$, and three different viscosity ratios, $\nu_0/\nu_\infty = 1.3$, $\nu_0/\nu_\infty = 1.4$, and $\nu_0/\nu_\infty = 1.5$. 
As in Fig.~3 of the main text, we observe a continuous decrease of the area fraction of turbulent regions with decreasing activity $a$ until a critical point is reached at $a=a_4^\ast$. There, the whole system becomes macroscopically quiescent.
We observe that an increase in the viscosity ratio $\nu_0/\nu_\infty$ shifts the critical activity $a_4^\ast$ to higher values.
The inset of Supplementary Figure~\ref{fig:transitionSuppl} shows the scaling of $\Phi$ when approaching the critical activity  from above.
As in Fig.~3 of the main text, we again observe power-law scaling $\Phi \propto (a - a_4^\ast)^{\beta^\ast}$, although
the exponent $\beta^\ast$ is different.
Via fitting to the curves, we obtain $\beta^\ast = 0.103(5)$, where the number in brackets denotes an error estimate corresponding to a confidence interval of $\SI{95}{\percent}$.
As Supplementary Figure~\ref{fig:transitionSuppl} shows, $\beta^\ast$ does not seem to depend on the value of the viscosity ratio $\nu_0/\nu_\infty$.
These results suggest that the exponent is independent or only weakly dependent on both $\nu_0/\nu_\infty$ and $n$, while it depends on the value of the crossover shear rate $\zeta$.

\begin{figure}
\includegraphics[width=0.61\linewidth]{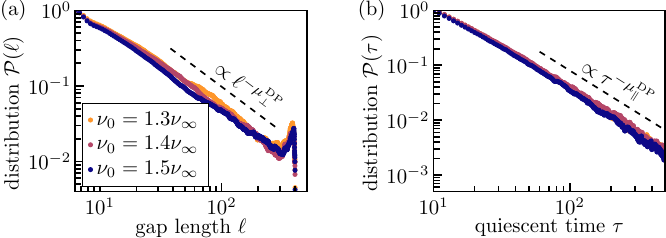}
\caption{\label{fig:DPSuppl}Features of the spatial and temporal structures very close to the critical point in terms of the distributions $\mathcal{P}$ of (a) the distances (lengths of the gaps) between the turbulent regions $\ell$ and (b) the time intervals of local quiescence $\tau$.
The distributions are normalized by the value at smallest $\ell$ or $\tau$.
Parameter values are set to $\zeta = 1$, $n=4$, and the activity is set to $a=1.19399$ for $\nu_0 = 1.3 \nu_\infty$, $a=1.22216$ for $\nu_0 = 1.4 \nu_\infty$, and $a=1.24417$ for $\nu_0 = 1.5 \nu_\infty$.
The activities are chosen as close to the critical points as possible, compare Supplementary Figure~\ref{fig:transitionSuppl}.
An emergent power-law scaling near the transition for elevated $\ell$ and $\tau$ is consistent with $2+1$ directed percolation (``DP''), as indicated by the dashed lines, which represent the exponents of directed percolation $\mu_\perp^\mathrm{DP} = 1.204(2)$ and $\mu_\parallel^\mathrm{DP} = 1.5495(10)$~\cite{takeuchi2007directed,hinrichsen2000non}.}
\end{figure}

As discussed in the main text, recent studies~\cite{takeuchi2007directed,sipos2011directed,lemoult2016directed,sano2016universal,doostmohammadi2017onset} have shown that the emergence of turbulence in various systems can be understood in terms of a transition of directed percolation (``DP'').
Whether the transition in the shear-thinning active fluid can be linked to directed percolation as well is thus a natural question. 
However, the fact that the exponent $\beta^\ast$ does not obtain a universal value for different sets of parameters complicates the picture.
It follows that the activity apparently does not directly correspond to the control parameter in directed percolation.
Nevertheless, we can establish a potential link by exclusively focusing on the behavior at the critical point and not on how it is approached.

In directed percolation, the spatial and temporal correlations become scale-free at the critical point.
As motivated by Refs.~\citenum{lemoult2016directed,takeuchi2007directed,chantry2017universal}, and as discussed in the main text, such correlations can be characterized by the distributions $\mathcal{P}(\ell)$ and $\mathcal{P}(\tau)$, where $\ell$ denotes the gap lengths between turbulent domains and $\tau$ the time intervals of quiescence.
As shown in the main text for $\zeta = 2$, $\nu_0/\nu_\infty=1.5$, and various values of $n$, we observe power-law scaling, $\mathcal{P}(\ell)\propto\ell^{-\mu_\perp}$ and $\mathcal{P}(\tau)\propto\tau^{-\mu_\parallel}$ very close to the critical point, reflecting self-similarity.
The obtained results are consistent with the exponents $\mu_\perp^\mathrm{DP}$ and $\mu_\parallel^\mathrm{DP}$ of $2+1$ directed percolation.

To complement the results presented in the main text, we performed additional simulations regarding the nature of the transition for the parameter sets discussed above, that is, $\zeta = 1$, $n=4$, and the viscosity ratios, $\nu_0/\nu_\infty = 1.3$, $\nu_0/\nu_\infty = 1.4$, and $\nu_0/\nu_\infty = 1.5$.
Here, we set the activities to values as close to the critical point as possible, $a=1.19399$ for $\nu_0/\nu_\infty = 1.3$, $a=1.22216$ for $\nu_0/\nu_\infty = 1.4$, and $a=1.24417$ for $\nu_0/\nu_\infty = 1.5$.
The results for the distributions $\mathcal{P}(\ell)$ and $\mathcal{P}(\tau)$ are shown in Supplementary Figure~\ref{fig:DPSuppl}.
As in Fig.~3 in the main text, we observe power-law scaling that is consistent with the $2+1$ directed percolation exponents, denoted by the dashed lines.
The exponents determined via fitting to the data points are close to the exponents expected for $2+1$ directed percolation as well, see Table~1 in the main text.

\section{Impact of initial conditions}

\begin{figure}
\includegraphics[width=0.96\linewidth]{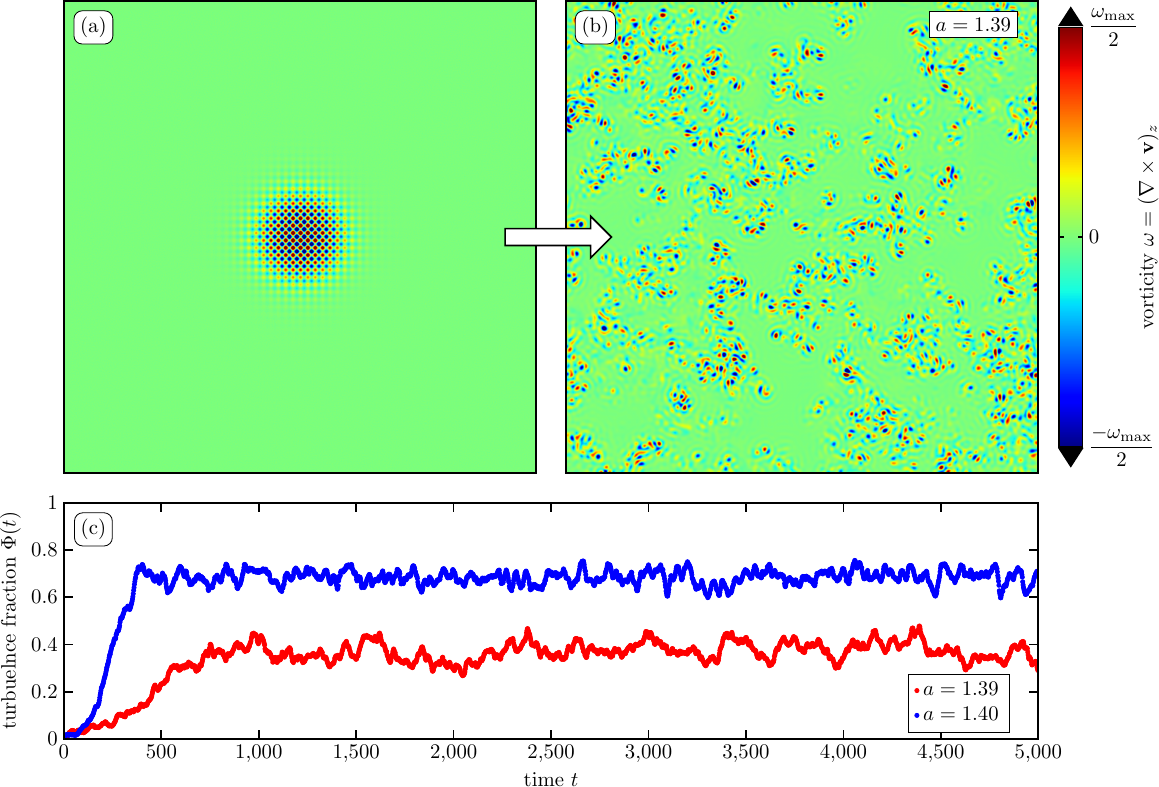}
\caption{\label{fig:testInitialValues}Development of patchy turbulence from a localized lattice-like vortical flow pattern. (a) Vorticity field at the beginning of the simulation. We start from a quiescent state with a localized perturbation in the center of the system consisting of a regular lattice of vortices. (b) Snapshot of the vorticity field that develops from the conditions in (a) at an activity of $a=1.39 < a_\mathrm{th}$. The local perturbations lead to self-sustained patchy turbulence. (c) Evolution of the turbulence fraction $\Phi(t)$ starting from the conditions in (a) for activities of $a=1.39$ and $a=1.4$, both below $a_\mathrm{th}$. It takes $500$ to $1000$ time units for $\Phi(t)$ to acquire its final fluctuating but statistically steady state. There, $\Phi(t)$ fluctuates around a constant value.}
\end{figure}

In the main text, we demonstrate the hysteretic behavior observed when varying the strength of activity $a$.
Increasing $a$ from small values and starting from a quiescent state, $\mathbf{v}(\mathbf{x},t) = \mathbf{0}$, the system only develops turbulence once $a$ exceeds the instability threshold $a_\mathrm{th}$.
Contrarily, starting with a turbulent state at $a > a_\mathrm{th}$ and decreasing the activity, turbulence is sustained below the threshold.
In this hysteretic regime of turbulence, locally high shear rates associated with turbulence lead to smaller viscosity and, thus, enable the turbulent domains to persist.
In the following, we check for possible dependencies of such dynamic scenarios on the initial conditions of the simulations.
Instead of slowly decreasing the activity into the hysteretic regime, we set $a < a_\mathrm{th}$ from the beginning of the simulation.
The system is initialized with a non-vanishing flow pattern in the center of the system surrounded by a large quiescent area.
For this central initial core of flow, we use a lattice of vortices according to
\begin{equation}
\label{eq:lattice}
\begin{aligned}
v_x(x,y) = A \sin(y) g(x,y), \\
v_y(x,y) = A \sin(x) g(x,y),
\end{aligned}
\end{equation}
where $A$ is the amplitude.
The function $g(x,y)$ is employed to limit the presence of this vortex lattice to the center of the system. 
We here use
\begin{equation}
\label{eq:localization}
g(x,y) = \{1 + \tanh[(s/2 - |x|)/\delta]\}\{1 + \tanh[(s/2 - |y|)/\delta]\}/4, 
\end{equation}
where $s$ is the size of the lattice and $\delta$ quantifies the thickness of the transition region between the vortex lattice and its quiescent surroundings.

Supplementary Figure~\ref{fig:testInitialValues} illustrates how patchy turbulence develops from these initial conditions.
There, we set the system size to $128\pi \times 128\pi$, the amplitude to $A=2$, and the size of the lattice and transition region to $s = 16\pi$ and $\delta = 6.4 \pi$.
As before, we add small initial random perturbations to each velocity component at every grid point, following a uniform distribution over $[-0.01,0.01]$.
The resulting vorticity field is shown in Supplementary Figure~\ref{fig:testInitialValues}(a).
The other parameters are set to $\nu_0/\nu_\infty = 1.5$, $\zeta = 2$, $n=4$, and $a = 1.39 < a_\mathrm{th}$. 
After $1000$ time units, turbulence has spread throughout the whole system, see Supplementary Figure~\ref{fig:testInitialValues}(b).
To explain this behavior, we remark that the initial lattice-like flow state reduces the local viscosity. Consequently, vortices are maintained and the area of turbulence grows. 
Quantifying the spatiotemporal evolution, we plot the turbulence fraction $\Phi$ as a function of time $t$ in Supplementary Figure~\ref{fig:testInitialValues}(c) at $a=1.39$ and $a=1.4$, both below the threshold value $a_\mathrm{th}$.
After an initial growth, $\Phi(t)$ fluctuates around a constant value, which marks the appearance of a statistically stationary state.
Here, the time average of $\Phi$ depends on activity.
Importantly, despite the modified initialization, we recover the values obtained from the simulations as reported in Fig.~2(b) of the main manuscript, $\Phi \approx 0.38$ for $a=1.39$ and $\Phi \approx 0.68$ for $a=1.4$.
We further explored the behavior for smaller amplitudes of the initial vortex pattern, for example, $A = 0.5$.
Here, the initial lattice-like flow pattern is too weak and decays over time, leading to a fully quiescent state.

Overall, these results demonstrate that the exact initial conditions of the simulations are not decisive for the statistical appearance of the resulting state, as long as they are capable of inducing the turbulent state. 
That is, if the initial flow patterns are strong enough, turbulent patches will develop locally and over time spread through the whole system.
During this process, the memory of the initial conditions is lost, concerning the overall statistical appearance of the system.
If the initial flow patterns are too weak in the hysteretic regime, then turbulent states do not develop and the system turns back to quiescence. 
The only states observed are thus either complete quiescence, $\mathbf{v}(\mathbf{x},t) = \mathbf{0}$, or turbulent states.

\section{Robustness of the results}

In order to check for finite-size effects, we plot the turbulence fraction $\Phi$ as a function of activity $a$ for different system sizes at $\zeta = 1$, $n=4$, and $\nu_0/\nu_\infty = 1.4$ in Supplementary Figure~\ref{fig:transitionSize}.
Clearly, the data points for system sizes of $64\pi\times 64\pi$ and $128\pi\times 128\pi$ coincide, indicating the robustness of the results.
In particular, the systems considered in our investigations are large enough so that finite-size effects do not play any obvious role for these data. \\

\begin{figure}[h]
\includegraphics[width=0.56\linewidth]{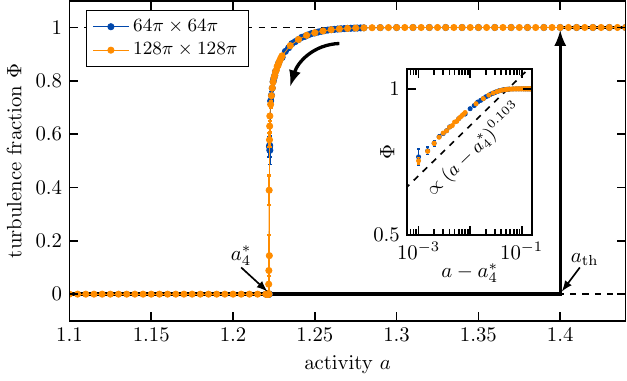}
\caption{\label{fig:transitionSize}Time-averaged area fraction of turbulent domains $\Phi$ as a function of activity $a$ for a crossover shear rate $\zeta = 1$ and degree of sharpness of the crossover $n=4$ of shear thinning.
When changing the system sizes as indicated for $\nu_0 / \nu_\infty = 1.4$, we do not observe any significant change in the area fraction of turbulent regions as a function of activity $a$.}
\end{figure}

%\bibliography{references}
